\documentclass[twocolumn]{aastex63}
\usepackage{amsmath,apjfonts,amsfonts,amssymb,enumerate,graphicx, longtable, float, times,textcomp,verbatim,url,nicefrac}
\usepackage[utf8]{inputenc}

\def\aj{AJ}%
\def\apj{ApJ}%
\def\apjl{ApJ}%
\def\apjs{ApJS}%
\def\ao{Appl.~Opt.}%
\def\aap{A\&A}%
\def\mnras{MNRAS}%
\newcommand\mki{Department of Physics and Kavli Institute for Astrophysics and Space Research, Massachusetts Institute of Technology, Cambridge, MA 02139, USA}
\newcommand\cfa{Center for Astrophysics \textbar \ Harvard \& Smithsonian, 60 Garden Street, Cambridge, MA 02138, USA}
\newcommand\gsfc{NASA's Goddard Space Flight Center, Greenbelt, MD 20771, USA}

\tabletypesize{\footnotesize}
\shorttitle{TESS Target Selection}
 \shortauthors{Fausnaugh et al.}

\begin{document}

\title{The TESS Mission Target Selection Procedure}

\author[0000-0002-9113-7162]{Michael~Fausnaugh}
\affil{\mki}

\author{Ed~Morgan}
\affiliation{\mki}

\author[0000-0001-6763-6562]{Roland\ Vanderspek}
\affil{\mki}

\author[0000-0002-3827-8417]{Joshua Pepper}
\affiliation{Lehigh University, Department of Physics, 16 Memorial Drive East, Bethlehem, PA, 18015, USA}


\author[0000-0002-7754-9486]{Christopher~J.~Burke}
\affil{\mki}

\author[0000-0001-8172-0453]{Alan~M.~Levine}
\affiliation{\mki}

\author{Alexander~Rudat}
\affil{\mki}

\author{Jesus~Noel~S.~Villase\~{n}or}
\affil{\mki}

\author{Michael~Vezie}
\affiliation{\mki}

\author{Robert~F.~Goeke}
\affiliation{\mki}

\author{George R.\ Ricker}
\affil{\mki}

\author[0000-0001-9911-7388]{David W.\ Latham}
\affil{\cfa}

\author[0000-0002-6892-6948]{S.~Seager}
\affil{\mki}
\affil{Department of Earth, Atmospheric, and Planetary Sciences, Massachusetts Institute of Technology, Cambridge, MA 02139, USA}
\affiliation{Department of Aeronautics and Astronautics, MIT, 77 Massachusetts Avenue, Cambridge, MA\
 02139, USA}

\author[0000-0002-4265-047X]{Joshua~N.~Winn}
\affil{Department of Astrophysical Sciences, Princeton University, 4 Ivy Lane, Princeton, NJ 08544, USA}

\author[0000-0002-4715-9460]{Jon M.\ Jenkins}
\affil{NASA Ames Research Center, Moffett Field, CA, 94035, USA}


\author[0000-0001-7204-6727]{G.~\'A.~Bakos}
\affiliation{Department of Astrophysical Sciences, Princeton University,  4 Ivy Lane,  Princeton, NJ 08544, USA}
\affiliation{Institute for Advanced Study, 1 Einstein drive, Princeton, NJ 08540}

\author[0000-0001-7139-2724]{Thomas Barclay} 
\affiliation{\gsfc}
\affiliation{University of Maryland, Baltimore County, 1000 Hilltop Cir, Baltimore, MD 21250, USA}

\author[0000-0002-3321-4924]{Zachory~K.~Berta-Thompson}
\affiliation{University of Colorado Boulder, Boulder, CO 80309, USA}

\author[0000-0002-0514-5538]{Luke G.~Bouma}
\affiliation{Department of Astrophysical Sciences, Princeton University, 4 Ivy Lane, Princeton, NJ 08544, USA}

\author[0000-0003-2519-3251]{Patricia T.~Boyd}
\affiliation{\gsfc}

\author[0000-0002-9314-960X]{C.~E.~Brasseur}
\affiliation{Space Telescope Science Institute, 3700 San Martin Drive, Baltimore, MD 21218, USA}

\author[0000-0002-0040-6815]{Jennifer~Burt}
\affiliation{Jet Propulsion Laboratory, California Institute of Technology, 4800 Oak Grove Drive, Pasadena, CA 91109, USA}

\author[0000-0003-1963-9616]{Douglas A. Caldwell}
\affiliation{NASA Ames Research Center, Moffett Field, CA, 94035, USA}
\affiliation{SETI Institute, Mountain View, CA 94043, USA} 

\author[0000-0002-9003-484X]{David Charbonneau}
\affiliation{\cfa}

\author[0000-0001-5137-0966]{J.~Christensen-Dalsgaard}
\affiliation{Stellar Astrophysics Centre, Department of Physics and Astronomy, Aarhus University, Ny Munkegade 120, DK-8000 Aarhus C, Denmark}

\author[0000-0003-4003-8348]{Mark Clampin}
\affiliation{\gsfc}

\author[0000-0001-6588-9574]{Karen~A.~Collins}
\affiliation{\cfa}

\author[0000-0001-8020-7121]{Knicole D. Col\'{o}n}
\affiliation{\gsfc}

\author[0000-0002-3657-0705]{Nathan De Lee}
\affiliation{Department of Physics, Geology, and Engineering Technology, Northern Kentucky University, Highland Heights, KY 41099, USA}
\affiliation{Vanderbilt University, Department of Physics \& Astronomy, 6301 Stevenson Center Lane, Nashville, TN 37235, USA}

\author{Edward~Dunham}
\affiliation{Lowell Observatory, 1400 W. Mars Hill Road, Flagstaff, AZ 86001}

\author[0000-0003-0556-027X]{Scott W.~Fleming}
\affiliation{Space Telescope Science Institute, 3700 San Martin Drive, Baltimore, MD 21218, USA}

\author[0000-0003-0241-2757]{William Fong}
\affiliation{\mki}
\author[0000-0001-9828-3229]{Aylin Garcia Soto}
\affiliation{\mki}
\affiliation{Department of Physics and Astronomy, Dartmouth College Hanover, NH 03755}    

\author[0000-0003-0395-9869]{B. Scott Gaudi}
\affiliation{Department of Astronomy, The Ohio State University, 140 W. 18th Avenue, Columbus, OH 43210, USA}

\author[0000-0002-5169-9427]{Natalia M.~Guerrero}
\affiliation{\mki}

\author[0000-0002-2135-9018]{Katharine Hesse}
\affiliation{\mki}

\author[0000-0002-1139-4880]{Matthew J.~Holman}
\affiliation{\cfa}

\author[0000-0003-0918-7484]{Chelsea X.\ Huang}
\altaffiliation{Juan Carlos Torres Fellow}
\affiliation{\mki}

\author[0000-0002-0436-1802]{Lisa Kaltenegger}
\affiliation{Carl Sagan Institute, Cornell University, 302 Space Science Building, Ithaca, NY 14850,  USA}

\author[0000-0001-6513-1659]{Jack J. Lissauer}
\affiliation{NASA Ames Research Center, Moffett Field, CA, 94035, USA}

\author{Scott McDermott}
\affiliation{Proto-Logic LLC, 1718 Euclid Street NW, Washington, DC 20009, USA}

\author{Brian McLean}
\affiliation{Space Telescope Science Institute, 3700 San Martin Drive, Baltimore, MD 21218, USA}

\author[0000-0002-4510-2268]{Ismael Mireles}
\affiliation{\mki}
\affiliation{Department of Physics and Astronomy, University of New Mexico, 210 Yale Blvd NE, Albuquerque, NM 87106, USA}

\author[0000-0001-7106-4683]{Susan E.~Mullally}
\affiliation{Space Telescope Science Institute, 3700 San Martin Drive, Baltimore, MD 21218, USA}

\author[0000-0002-0582-1751]{Ryan J. Oelkers}
\affiliation{Vanderbilt University, Department of Physics \& Astronomy, 6301 Stevenson Center Lane, Nashville, TN 37235, USA}

\author[0000-0001-8120-7457]{Martin Paegert}
\affiliation{\cfa}

\author[0000-0001-5449-2467]{Andr\'as P\'al}
\affiliation{Konkoly Observatory, Research Centre for Astronomy and Earth Sciences, H-1121 Budapest, Konkoly Thege Mikl\'os \'ut 15-17, Hungary}

\author[0000-0003-1309-2904]{Elisa V.~Quintana}
\affiliation{\gsfc}

\author[0000-0003-2519-3251]{S. A. Rinehart}
\affiliation{\gsfc}
\affiliation{NASA Headquarters, Planetary Science Division, Washington DC 20546, USA}

\author[0000-0003-1286-5231]{David R.~Rodriguez}
\affiliation{Space Telescope Science Institute, 3700 San Martin Drive, Baltimore, MD 21218, USA}

\author[0000-0003-4724-745X]{Mark Rose}
\affiliation{NASA Ames Research Center, Moffett Field, CA, 94035, USA}

\author[0000-0001-7014-1771]{Dimitar D. Sasselov}
\affiliation{\cfa}

\author{Joshua E. Schlieder}
\affiliation{\gsfc}

\author[0000-0001-5401-8079]{Lizhou Sha}
\affiliation{\mki}
\affiliation{Department of Astronomy, University of Wisconsin, Madison, WI 53706-1507, USA}

\author[0000-0002-1836-3120]{Avi~Shporer}
\affiliation{\mki}

\author[0000-0002-6148-7903]{Jeffrey C. Smith}
\affiliation{NASA Ames Research Center, Moffett Field, CA, 94035, USA}
\affiliation{SETI Institute, Mountain View, CA 94043, USA} 

\author[0000-0002-3481-9052]{Keivan G. Stassun}
\affiliation{Vanderbilt University, Department of Physics \& Astronomy, 6301 Stevenson Center Lane, Nashville, TN 37235, USA}
\affiliation{Fisk University, Department of Physics, 1000 18th Avenue N., Nashville, TN 37208, USA}

\author[0000-0002-1949-4720]{Peter Tenenbaum}
\affiliation{NASA Ames Research Center, Moffett Field, CA, 94035, USA}
\affiliation{SETI Institute, Mountain View, CA 94043, USA} 

\author[0000-0002-8219-9505]{Eric~B.~Ting}
\affiliation{NASA Ames Research Center, Moffett Field, CA, 94035, USA}

\author[0000-0002-5286-0251]{Guillermo Torres}
\affiliation{\cfa}

\author[0000-0002-6778-7552]{Joseph D. Twicken}
\affiliation{NASA Ames Research Center, Moffett Field, CA, 94035, USA}
\affiliation{SETI Institute, Mountain View, CA 94043, USA}

\author[0000-0001-7246-5438]{Andrew~Vanderburg}
\affiliation{Department of Astronomy, University of Wisconsin, Madison, WI 53706-1507, USA}

\author[0000-0002-5402-9613]{Bill~Wohler}
\affiliation{NASA Ames Research Center, Moffett Field, CA, 94035, USA}
\affiliation{SETI Institute, Mountain View, CA 94043, USA}

\author[0000-0003-1667-5427]{Liang Yu}
\affiliation{\mki}

\begin{abstract}
    We describe the target selection procedure by which stars are selected for 2-minute and 20-second observations by TESS.  We first list the technical requirements of the TESS instrument and ground systems processing that limit the total number of target slots.  We then describe algorithms used by the TESS Payload Operation Center (POC) to merge candidate targets requested by the various TESS mission elements (the Target Selection Working Group, TESS Asteroseismic Science Consortium, and Guest Investigator office). Lastly, we summarize the properties of the observed TESS targets over the two-year primary TESS mission.  We find that the POC target selection algorithm results in 2.1 to 3.4 times as many observed targets as target slots allocated for each mission element.  We also find that the sky distribution of observed targets is  different from the sky distributions of candidate targets due to technical constraints that require a relatively even distribution of targets across the TESS fields of view.  We caution researchers exploring statistical analyses of TESS planet-host stars that the population of observed targets cannot be characterized by any simple set of criteria applied to  the properties of the input Candidate Target Lists.
\end{abstract}
\keywords{TESS}

\section{Introduction \label{sec:intro}}
The Transiting Exoplanet Survey Satellite (TESS, \citealt{Ricker2015}) completed its two-year primary mission in July of 2020, and is currently running a 26 month extended mission.  The primary goal of TESS is to identify small exoplanets around bright stars.  To accomplish this, TESS observes 20\,000 targets nearly continuously for approximately 27 days at a time, searching for subtle signatures of transiting planets.

A key operational aspect of the TESS mission is that subarrays of pixels associated with the 20\,000 target stars are saved at two-minute cadence (hereinafter "2-minute cadence"), in order to resolve the transit profiles of small, short-period exoplanets.  However, there is a finite number of such subarrays that can be saved in each observing sector due to limits on the instrument's data storage capacity and requirements on data processing throughput in ground systems.  Given the limited number of target slots, it is imperative to identify the best target stars for the mission's transiting planet search in advance.  To prepare for TESS observations, the TESS Target Selection Working Group (TSWG) constructed a list of suitable stars and an associated priority for observation by TESS, the so called "candidate target list" (CTL).  Details of the construction of the CTL and the all-sky TESS Input Catalog (TIC) can be found in \citet{Stassun2018, Stassun2019}.  The CTL covers the entire sky and constitutes the parent sample of the 20\,000 stars selected for 2-minute observations in each TESS observing sector.

Continuous 2-minute monitoring by TESS also opens new scientific opportunities in the fields of asteroseismology (e.g., \citealt{Antoci2019, Huber2019, Pedersen2019, Schofield2019, Metcalfe2020}), stellar variability (\citealt{Hodapp2019, Balona2020, Burssens2020, Nazel2020, Tajiri2020, Plachy2021}),  solar system science (\citealt{Zieba2019, Farnham2019}) and other areas of astrophysical research (\citealt{ Bell2019, Littlefield2019, Wang2020}).  To capitalize on these opportunities, additional high priority targets for the TESS mission were  identified by the TESS Asteroseismic Science Consortium (TASC) and a call to the community through the TESS Guest Investigator (GI) office.\footnote{The GI office announces a call for proposals approximately once per year for the upcoming cycle of TESS observations.}  In the same way as the TSWG, the TASC and GI office each produced a CTL consisting of a list of desired targets and an associated priority for observations. 


In the extended mission, TESS added a new twenty-second cadence data mode (hereinafter "20-second cadence").  The number of target slots is more limited for 20-second cadence observations than for 2-minute observations, and a similar target selection scheme is required for 20-second observations.  In 2020, the GI office  produced an additional CTL specifying prioritized targets for 20-second cadence observations.  The 20-second cadence targets also influence the 2-minute target selection, because the mission requires that all 20-second cadence targets also be saved at 2-minute cadence and therefore uses 1\,000 of the 20\,000 target slots.

The initial versions of the CTLs from the TSWG, TASC, and GI office were delivered to  the TESS Payload Operation Center (POC) before TESS was launched in 2018 April.  The CTLs have been occasionally revised since then; in particular, the GI  CTL is updated after every proposal cycle with new targets selected by peer review.  In a similarly way,  20-second cadence CTLs were delivered in the summer of 2020 before the start of the extended mission, and will be updated after every GI proposal cycle.

For every TESS observing sector, the POC selects targets for observations from the most recent CTLs, identifies the associated pixel subarrays, and produces data tables and command scripts that are uploaded to the spacecraft.  The POC target selection must maximize the science return of the mission by selecting high priority targets from each CTL, while fulfilling the technical requirements imposed by the limited number of target slots.   

In this paper, we describe the target selection process used by the POC to select target stars for 2-minute and 20-second cadence observations.  In \S\ref{sec:technical_constraints}, we enumerate the technical requirements of the mission that constrain target selection.  In section \S\ref{sec:target_selection}, we describe the algorithms that the POC uses to select targets from each CTL.  In \S\ref{sec:results} we  assess the sample of stars  that were selected for observations in the primary mission, and we summarize the results in \S\ref{sec:conclusion}.

\floattable
\begin{deluxetable}{lll}
\tablewidth{0pt}
\tablecaption{Technical Constraints on Target Selection\label{tab:constraints}}
\tablehead{
 \colhead{Locus} & \colhead{Purpose} &\colhead{Constraint}
 }
 \startdata
 SSR Storage & Limited data storage volume on instrument & $<$50\% data storage capacity per sector \\
 Photometer Performance Assessment & Monitor instrument performance & $\leq$120 PPA stars per CCD\\
Astrometric Registration & Define WCSs for every image & $\geq$100 PPA stars per CCD, and evenly spaced \\
 Presearch Data Conditioning & Remove systematic errors from mission supplied light curves & As many targets as possible per CCD\\
 Transiting Planet Search & Return mission supplied planet candidates  within two months& $\leq$20\,000 targets per sector and\\
 &\phantom{aaaa}  of data downlink & \phantom{aaaa}$\leq$2\,000 targets per CCD\\
\enddata
\end{deluxetable}

\section{Technical Constraints on Target Selection\label{sec:technical_constraints}}
The TESS instrument consists of four shutterless cameras, each with a 24x24 square degree field of view, and four CCD detectors with a pixel scale of approximately 21 arcseconds per pixel.  A Data Handling Unit (DHU) computer converts the stream of images from the CCDs to digital files and stores the files until they are downloaded from the spacecraft.  The CCDs are continuously read out every two seconds, and the DHU processes all 16 CCDs in parallel.  Subarrays of pixels from each CCD are stacked and saved every 20 seconds and two minutes, while the full frame is saved every 10 minutes (previously every 30 minutes in the primary mission from 2018 July to 2020 July).\footnote{The DHU also mitigates against cosmic ray hits by discarding the high and low values of pixels in blocks of 10 two-second exposures \citep{TIH}.  The cosmic ray rejection is applied to the 2-minute subarrays and FFIs, but not to the 20-second cadence subarrays.}  

The full flight image processing flow and DHU details can be found in \citet{TIH}.  For the purpose of target selection, the DHU includes a Flash Memory Card that serves as a Solid State Recorder (SSR) and can hold 192 GB of science data files.  The data files are compressed with a 27-bit Huffman table encoding scheme,\footnote{In the primary mission, a 24-bit Huffman table was used.} which achieves an average compression rate of about  3--8 bits/pixel, depending on the data mode and the level of background scattered light from earthshine and moonshine.  Even with this favorable compression factor, care must be taken that the total data volume of the 20 second, 2-minute, and full frame images (FFIs) does not exceed the storage capacity of the SSR.  In practice, the data volume limit is  $<$50\% of the SSR capacity in a single TESS orbit.  This requirement allows the DHU to save all data over two orbits, as a precaution against data downlink failures at the end of a single orbit.

The data volume limit implies a maximum number of pixels that can be saved at 20-second, 2-minute, and 10 minute (previously 30 minute) cadence.  Each 20-second and 2-minute target is assigned an 11x11 pixel subarray, and so the data volume limit also sets an upper limit on the number of targets that TESS can observe each sector.  However, both the duration of each TESS observing sector and the average compression factor over that time are variable (mainly due to changing scattered light backgrounds), so it is not possible to predict the total data volume as a function of the number of targets for each sector.  The mission therefore sets a maximum number of targets  and associated subarray pixels such that the total fill of the SSR will be close to but less than 50\% of its total capacity, even for longer TESS sectors and relatively poorer compression.

The target selection must also fulfill  processing requirements through the TESS ground systems, which include downlink at NASA's Deep Space Network (DSN), data reduction at the TESS Science Processing Operations Center (SPOC, \citealt{Jenkins2016}), and archiving/public release at the Mikulski Archive for Space Telescopes (MAST).  In particular, ground system processing must complete within two months of data downlink, and the SPOC data processing pipeline has the largest effect on the number of targets that can be analyzed in that time.  The SPOC is charged with 1) monitoring the instrument's performance, 2) defining World Coordinate Solutions (WCSs) for all pixel subarrays and FFIs, 3) providing systematic error corrected light curves for stars observed on subarrays, and 4) searching the light curves for transiting planet signatures and producing a list of candidates with associated diagnostic tests.  Each of these requirements affects the target selection:

\floattable
\begin{deluxetable}{ccclr}
\tablewidth{0pt}
\tablecaption{TESS Primary Mission Candidate Target Lists\label{tab:ctls}}
\tablehead{
\colhead{Order} &\colhead{Name} & \colhead{Origin} &\colhead{Content}
 &\colhead{N Slots}
 }
 \startdata
1 & ppa & POC & Engineering (PPA) targets. & 1\,920\\
2 & bright & POC & Stars with magnitude $\leq6$. &\\
3 & exo & TSWG & Stars most suitable for exoplanet discovery. & 13\,400\\
4 & astero & TASC & Stars suitable for asteroseismology observations. & 751\\
5 & GI  & GI Office & Targets selected through external proposals. & 1\,501\\
6 & DDT & PI/POC& Targets selected through the&1\,501 \\
 &      &        & Director's Discretionary Time program. & \\
\enddata
\end{deluxetable}

\begin{enumerate}
\item To monitor the instrument's performance, TESS observes a number of bright but unsaturated stars for each data mode.  These engineering targets are referred to as "photometer performance assessment" (PPA) stars.  PPA stars should be uncrowded and quiet (non-variable), in order to cleanly measure the instrument's response and stability.  A maximum of 120 PPA stars are allowed per CCD, which provides a small margin above the minimum number of PPA stars needed for accurate WCSs.

\item To define WCSs, the SPOC measures the positions of bright, uncrowded stars and fits for transformations  between celestial and detector coordinates.  PPA stars are ideally suited to this purpose, but must be well distributed  across the TESS fields of view in order to constrain the WCS for each CCD. The accuracy of the WCSs is determined by the number of PPA stars and their distribution in the focal plane.  A minimum of 100 PPA stars is required per CCD for this purpose.

\item To provide systematic error correction for TESS light curves, the SPOC uses the Presearch Data Conditioning (PDC) algorithm developed for the \textit{Kepler} mission.  PDC uses singular value decomposition (SVD) to find common trends in a large ensemble of light curves, and removes these trends using regularized fitting techniques.  PDC has demonstrated an excellent ability to preserve astrophysical variability on individual targets as long as there are enough light curves to identify common mode systematic trends \citep{Smith2020}. For TESS, systematic errors tend to vary from CCD to CCD, and so every CCD needs to have a large number of targets ($\gtrsim$1\,000) for PDC to effectively apply the systematic error corrections.

\item The TESS mission requires that data products and results of the transiting planet search be made public within two months of data downlink.  This requirement was set during mission development, based on the following considerations: (1) the pipeline processing must finish within 27 days so as to not fall behind real-time observations, (2) there must be margin for reprocessing if necessary, and (3) additional time is needed to review the pipeline results, export data products, and transfer data products to the public archive.  For the pessimistic scenario in which a full 27 days are needed to run the pipeline and the pipeline must be run twice, a limit of two months was set for the time between data downlink and public release.  To fulfill this requirement, the SPOC pipeline was tested against various throughput benchmarks.  The total computational cost primarily depends on the total number of targets submitted to the transiting planet search (TPS) and subsequently passed to data validation (DV) for fitting of planet candidate parameters and computation of diagnostic metrics.  Prelaunch tests showed that the pipeline can process 20\,000 targets per sector in a timely fashion,   with margin to reprocess if necessary.\footnote{Originally, this limit was 16\,000 source per sector, and was applied to  TESS Sectors~1--3.  In Sector~4, further testing allowed us to increase the limit to  20\,000.}  In order to make sure that the systematic error removal applied by PDC will complete on schedule, there is also a maximum of 2\,000 targets allowed per CCD.  Although these ground system requirements formally limit the total number of subarray targets that TESS observes each sector, the corresponding data volume is  close to the 50\% storage limit of the SSR after accounting for the FFIs.
\end{enumerate}

Table~\ref{tab:constraints} summarizes these 	constraints.  In brief, the technical aspects of the mission impose a maximum of 20\,000 2-minute targets with subarrays per sector, with a maximum of 2\,000 targets per CCD.  Each CCD must also have 100--120 stars selected as PPAs, in order to monitor the instrument performance and derive WCSs.  The PPA stars must be as bright as possible without saturating the detector and show little astrophysical variability, while being well-distributed across each CCD image. Finally, the POC attempts to maximize the number of targets per CCD, in order to improve the systematic error correction performed by PDC and better balance computational loads in the SPOC pipeline.  

\section{Target Selection \label{sec:target_selection}}

With these requirements in mind, we now turn to the algorithm that the POC uses to select targets.  Targets are drawn from the CTLs produced by the TSWG, TASC, and GI office.  Two more CTLs are produced by the POC:  a list of the very brightest stars in the sky (${\rm T_{mag}} \leq$ 6.0, where ${\rm T_{mag}}$ is the magnitude of the star for the TESS instrument response calibrated to Vega magnitudes), and a list of specially prioritized targets identified by the mission's Director's Discretionary Time (DDT) program.\footnote{\url{https://tess.mit.edu/science/ddt/}}  It was realized early in the mission that 2-minute observations of the brightest stars would have lasting scientific value at a small cost in terms of data volume.  The DDT program, on the other hand,  allows the community to obtain 2-minute or 20-second cadence TESS observations of interesting targets discovered after the yearly GI proposal cycle closes.

In total, there are therefore five CTLs from which to select targets:  the "exo" CTL produced by the TSWG; the "astero" CTL produced by the TASC; the "GI" CTL produced by the GI office; the "bright" CTL produced by the POC; and the "DDT" CTL also produced by the POC.  Throughout this paper we will use these labels (exo, astero, GI, bright, and DDT) to refer to each CTL.  Along with the list of PPA stars (described below), Table~\ref{tab:ctls} summarizes the labels, origins, and scientific purpose of each CTL.   

Every CTL contains a list of targets across the entire sky and an associated priority for observations.  The priorities are represented by a number between zero and unity, with higher numbers indicating a more preferred target.  Each CTL is constructed in isolation from the others, and so the priority rankings only represent relative preferences within a given CTL.  
\begin{itemize}
\item The exo CTL priorities are derived from the physical properties of stars, favoring dwarfs and smaller-radius stars.  Additional factors such as the star's brightness in the TESS bandpass, expected signal-to-noise ratio of transits, degree of blending, and the number of sectors for which a target is observable by TESS are also folded into the priority.  Full details are provided by \citet{Stassun2018,Stassun2019}.  The resulting priority distribution roughly follows a power law that mirrors the underlying stellar population.
\item Targets from the GI CTL are assigned based on peer-reviewed rankings of individual proposals, with minor adjustments made to ensure that the GI observing program has a balance of science topics and sizes of programs. Targets from a given science investigation are prioritized internally to that investigation, then targets from all proposed investigations are merged into a single list, ensuring that each target has a unique priority that varies between unity (highest priority) and zero (lowest priority). 
\item Targets from the astero CTL are assigned priorities based on the probability of detecting oscillations suitable for asteroseismology analysis.  For solar-like oscillations, the method is based on that of \citet{Chaplin2011}, and full details for the TESS astero CTL are provided in \citet{Schofield2019}.  Additional stars and priorities are added for eclipsing-binary timing standards \citep{vonEssen2020}, classical pulsators, and compact pulsators.
\item All targets in the bright CTL are assigned a priority of 1.0.
\item Targets from DDT proposals are assigned priorities by sequentially adding the most preferred target from each DDT program to a master DDT list.  The highest priority target from each DDT program is first added with a priority of 1.0.  The priority is then decremented by an amount calculated from the total number of DDT targets across all programs, and the next highest priority targets are added from each DDT program with the new priority.  This process is repeated until all DDT targets have been merged into a single list.
\end{itemize}
As we will see, the POC target selection method does not compare the priority distributions between different CTLs, and so the specific distribution of priorities within a given CTL does not matter as long as the relative ranking is correct.

In addition to these five CTLs, the POC generates a list of PPA stars for each sector.  This list effectively constitutes a sixth CTL that we label the "ppa CTL." However, the ppa CTL does not have an associated set of priorities because PPA stars are selected based on engineering requirements.   We describe the details of PPA star selection in \S\ref{sec:ppa_selection}, but otherwise this list of stars is treated as a sixth CTL in the following sections.

In \S\ref{sec:merging_ctls}, we describe how the POC merges prioritized targets from each CTL, and in \S\ref{sec:ccd_dist} we explain how the POC ensures that targets are well distributed across the CCDs.  Next, in \S\ref{sec:individual_targets} we discuss how individual targets from a given CTL are selected for observation and in \S\ref{sec:pixel_selection} how pixel subarrays are assigned to each target.  In \S\ref{sec:ppa_selection}, we detail the additional steps required for PPA star selection.  Throughout this section, we describe the details of the target selection for the primary mission (Sectors~1--26, 2018 July to 2020 July).  Changes to the target selection process and associated parameters for the extended mission and 20-second cadence targets (2020 July to 2022 September) are discussed in \S\ref{sec:em_selection}.

\subsection{Merging CTLs\label{sec:merging_ctls}}

\begin{figure}
  \centering
  \includegraphics[width=0.44\textwidth]{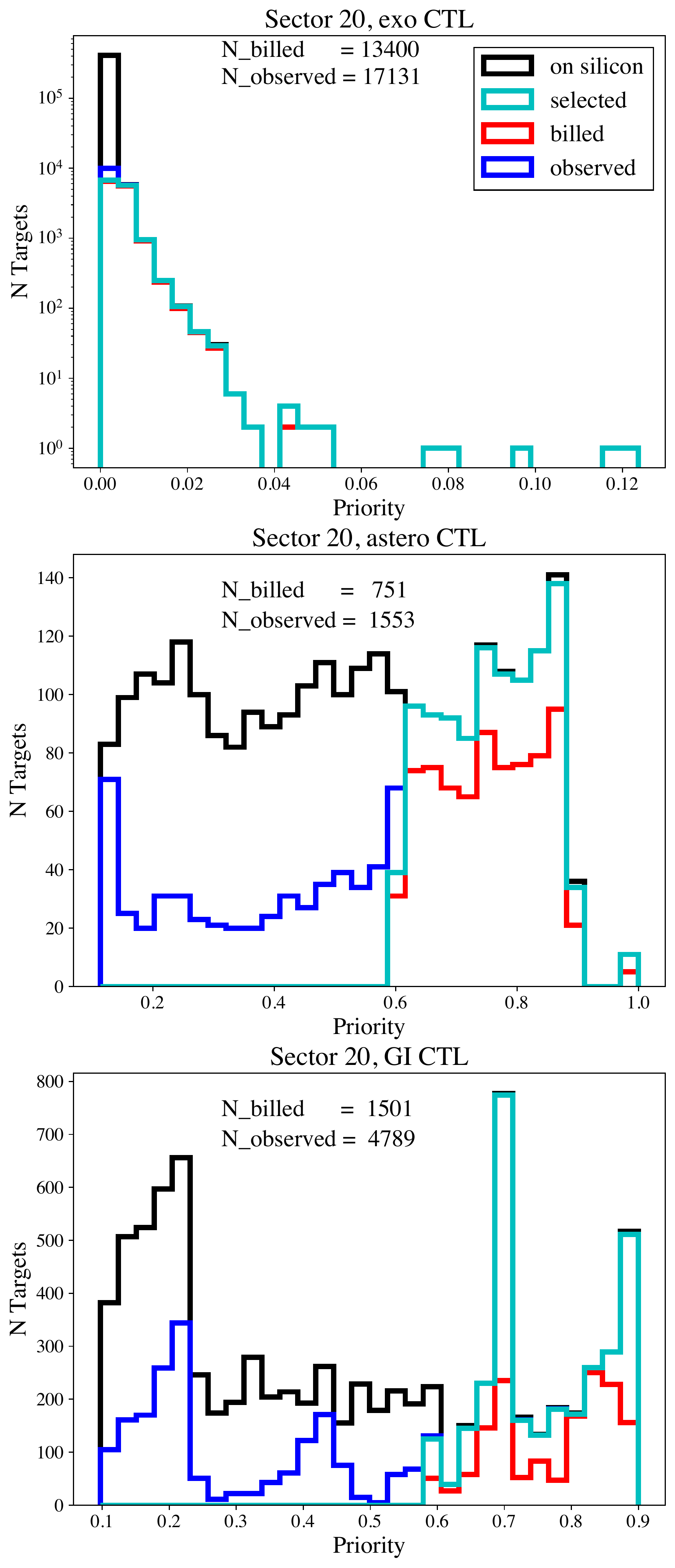}
  \caption{Example of target selection results for Sector~20.  Each panel shows the priority distribution for a different CTL (exo, astero, and GI---see \S\ref{sec:merging_ctls}).  The "on silicon" histogram shows the targets potentially observable in this sector and roughly follows the priority distribution of the full CTL.  The "selected" histogram shows the targets requested by a given CTL based on priority (see \S\ref{sec:individual_targets}) and added to the list of observed targets.   The "billed" histogram gives the targets counted against the target slots allocated by the POC (see Table~\ref{tab:ctls}).  The "observed" histogram gives the total objects observed at 2-minute cadence from a given CTL.  Note that the billed histogram is always a subset of the selected histogram, which is in turn a subset of the observed histogram.  The differences are due to overlap of targets in the various CTLs.  For the exo CTL, the majority of targets with priority greater than 0.005 were selected for observations. For the GI and astero CTLs, almost all of the high priority targets were observed due to overlap with the exo CTL.  See \S\ref{sec:merging_ctls} for more details.\label{fig:example_sector}}
\end{figure}

\begin{figure*}
  \centering
  \includegraphics[width=\textwidth]{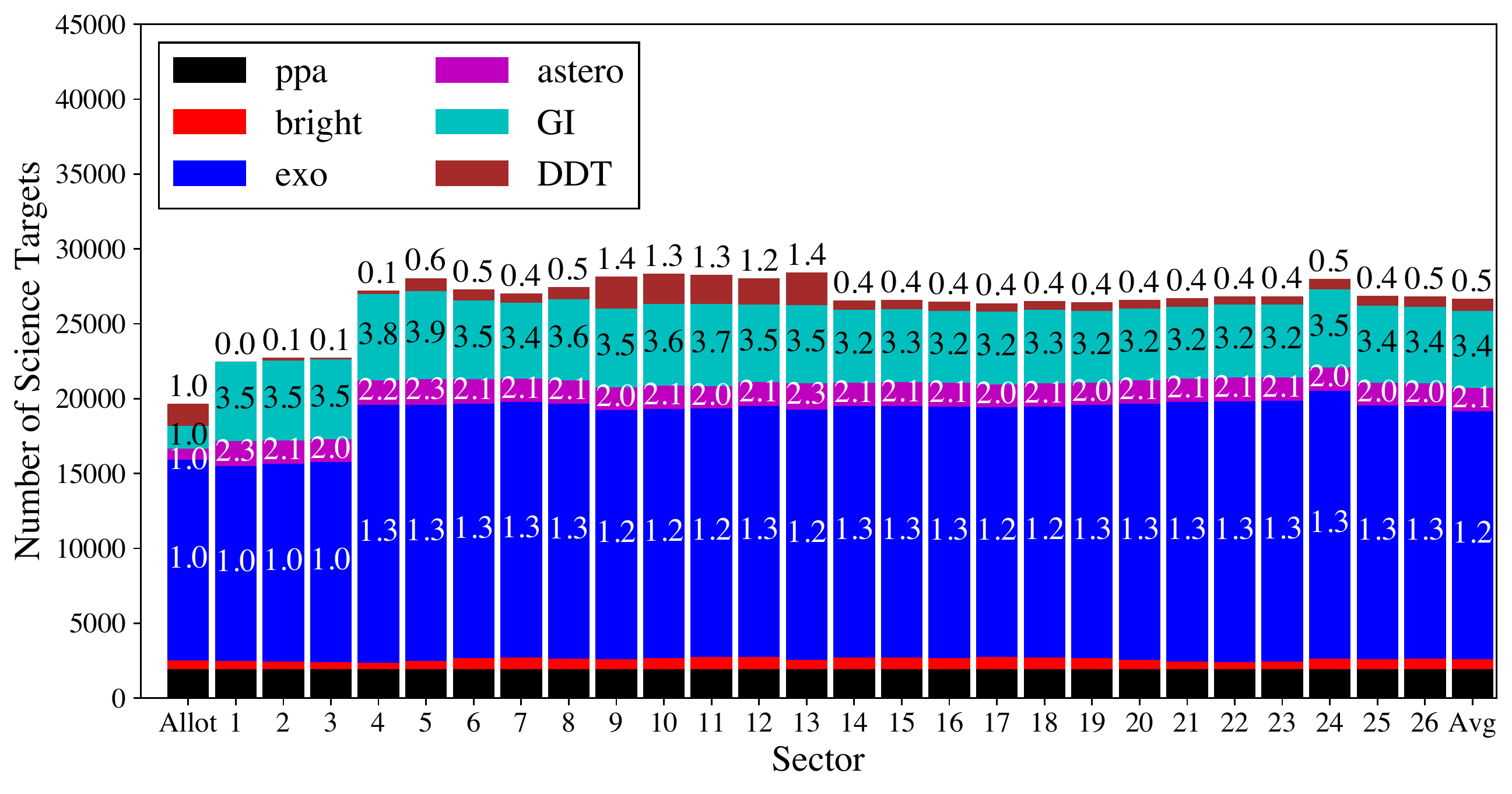}
  \caption{Number of observed targets from each CTL for each TESS observing sector in the primary mission.  The first bar gives the baseline allocation of targets per CTL defined by the POC to fulfill technical requirements (Table~\ref{tab:ctls}).  The last bar gives the average number of targets across all 26 sectors.  The numbers within/above each bar give the ratio of the number of targets observed to the allotment in the first bar---values greater than one demonstrate improved scientific return of the POC target selection algorithm compared to the baseline allotment.  While  the DDT target ratios are usually less than one, the total number of DDT targets is limited by the number of DDT candidate targets rather than the number of allocated target slots.   Also note that the first three sectors had a maximum of 16\,000 unique targets, while subsequent sectors had a maximum of 20\,000 unique targets.  Targets that appear in multiple CTLs are counted multiple times in this figure.\label{fig:sector_targs}}
\end{figure*}

The first challenge of the target selection process is to select targets from each CTL in a way  that  fairly allocates the limited number of target slots while maximizing the scientific return of the mission.  In order to do this, the POC adopted the following strategy.  

For any number of input CTLs, each CTL is assigned a maximum number of target slots.  The CTLs are then searched in a specific order for targets.  Once the maximum number of targets allocated to a CTL has been selected, the remaining targets from that CTL are ignored, and selection from the next CTL begins.     The number of targets for each CTL and the order in which CTLs are selected are adjustable parameters, allowing the POC to distribute the 20\,000 target slots across different CTLs in a flexible way.

For the TESS primary mission, targets were selected from the CTLs in the order shown in Table~\ref{tab:ctls}:  ppa, bright, exo, astero, GI, DDT.  The maximum number of target slots that were assigned to each CTL are also given in Table~\ref{tab:ctls}:  1\,920 for ppa, 13\,400 for exo, 751 for astero, 1\,501 for GI, and 1\,501 for DDT.  No limits were assigned to the bright CTL; in practice, the number of observable bright stars was between about 450 and 900 targets, all of which were assigned pixel subarrays.  These limits were initially based on predefined resource allocations.  However, the limits were refined over the first five TESS observing sectors, and we report the final  choices here.

The order in which the CTLs are searched for targets is important because a single target is often requested at high priority in multiple CTLs.  If a target appears in multiple CTLs, it is only counted against the total allotment in the first CTL that requested it. For example, a GI target that was first selected by the exo CTL is not counted against the 1\,501 slots allotted to the GI CTL, nor would an exo target that was first selected as a PPA star be counted against the 13\,400 slots for the exo CTL.  In this way, the POC algorithm does not need to compare the priorities between different CTLs.  However, there is  an advantage for a CTL to come up later in the target selection process, because a high priority target selected by an earlier CTL will not be counted against a later CTL's total allocation of targets.

The POC balances this advantage against the total number of assigned target slots---because the exo CTL is assigned nearly ten times as many target slots as each of the the astero, GI, and DDT CTLs, it obtains an overwhelming number of high priority targets regardless of its order in the target selection process.  Furthermore, the exo CTL would benefit less from the small number of targets selected from the astero, GI, and DDT CTLs than those CTLs benefit from the large number of selected exo CTL targets.  Selecting from the exo CTL first therefore increases the total number of targets observed from each CTL, and accordingly bolsters the total scientific return of the mission.  Similarly, the ppa and bright star lists represent required targets, and so are selected first to the benefit of the exo CTL.

Figure~\ref{fig:example_sector} shows an example of the target selection results for Sector~20, highlighting the extra targets observed from each CTL beyond the fiducial limits in Table~\ref{tab:ctls}.  Observable stars from the CTL ("on silicon") are shown with the black histograms.  "Selected" stars are stars that the CTL explicitly requested based on their priorities (see \S\ref{sec:individual_targets}) and are added to the observed target list.  However, each star is only "billed," that is counted against the target allotments in Table~\ref{tab:ctls}, if it has not already been selected by a previous CTL.  Because targets are selected from the exo CTL first, there are very few examples of this kind in the top panel of Figure~\ref{fig:example_sector} (i.e., the difference between the cyan and red histograms is very small).  The difference between the billed and selected histograms is more pronounced for the astero and GI CTLs because they have a high degree of overlap with the exo CTL.

Stars that are listed at low priority in a given CTL are also often selected at high priority in a different CTL.   The blue histograms in Figure~\ref{fig:example_sector} show the total "observed" stars, meaning all stars added to the target list that appear in the given CTL but were not necessarily added based on priority.  The lists of billed and selected stars are always a subset of the observed target list, but a substantial number of lower priority targets are also added.  These targets are freely observed for a given CTL in the sense that they are billed to a different CTL.  

Figure~\ref{fig:sector_targs} compares the number of these additional targets to the baseline allocation.  On average, the exo CTL receives about 25\% additional targets compared to the allocated slots, while the astero and GI receive a factor of about 2.1 and 3.4  times as many targets as allocated slots, respectively.  These extra targets  suggest an improved scientific return from the target selection algorithm compared to the baseline resource allocation given in Table~\ref{tab:ctls}.  In Appendix~A, we provide a table with the number of billed and observed targets for each sector from each CTL.

Finally, not all target star slots are filled for every CTL in every sector---this depends on the overall size of the CTLs and the distribution of CTL targets on the sky.  For example, the DDT CTL rarely reaches the full allotment of 1\,501 targets, especially for the first half of the primary mission when relatively few DDT proposals were submitted (see Appendix~A).  Thus, after selecting targets from all CTLs, any remaining target slots (up to 20\,000) are filled by re-selecting stars from the exo CTL. Typically, about 1\,500 additional exo targets are added in this way, so that the exo CTL usually reaches a total of $\sim$15\,000 observed targets per sector.

\begin{figure*}
  \centering
  \includegraphics[width=\textwidth]{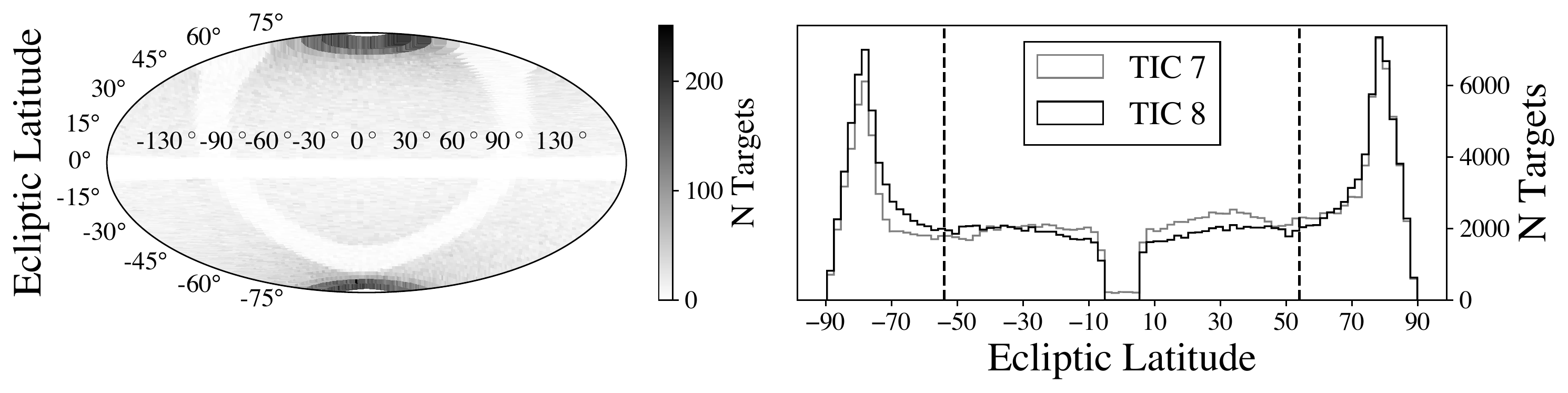}
  \caption{The left  panel shows the sky distribution in ecliptic coordinates of the top 200\,000 stars by priority in the exo CTL from TIC 8.  The bins subtend equal solid angles.  The curvilinear white band is the plane of the Galaxy, which was preferentially avoided because of the amount of crowding/contaminating flux expected in the large TESS pixels.  The right panel  shows the marginal distribution of the targets over ecliptic latitude (the TIC~7 CTL priorities from Year~1 are shown in gray for comparison).  Targets near the ecliptic poles can be observed for multiple sectors and  are assigned higher priority.  The vertical dashed lines show the approximate sky coverage of Cameras~3 and 4, which encompass about 55\% of the total available targets.  For a single sector, 90\% of the highest priority 20\,000 targets lie within these two cameras.  However, the light curve systematic error correction in the TESS data processing pipeline requires a more even distribution of targets across the TESS field of view.  We describe how the POC mitigates this issue in \S\ref{sec:ccd_dist}.\label{fig:top200k_ctl}}
\end{figure*}

\subsection{Distribution of Targets on TESS CCDs\label{sec:ccd_dist}}

In addition to the constraint of 20\,000 targets observed per sector, no more than 2\,000 targets can be selected per TESS CCD.  This requirement is usually in tension with the priorities of the CTLs, which  often assign higher priorities to targets near the ecliptic poles where TESS can observe the targets over multiple sectors.  Figure~\ref{fig:top200k_ctl} shows the sky distribution of the top 200\,000 targets by priority in the year~2 exo CTL---roughly 55\% of these targets are within 36 degrees of the ecliptic pole (the area of the sky observed by Cameras~3 and 4).  Of the highest priority 20\,000 observable targets per sector, on average about 60\% would be assigned to Camera~4 and 90\% would be assigned to Cameras~3 and 4 combined. 

A more even distribution of targets over the 16 TESS CCDs results in better systematic error correction by PDC, and the POC makes a concerted effort to more evenly distribute targets over the full TESS field of view.  To do this, the POC imposes a maximum number of targets from each CTL selected on each CCD.  The particular number is an adjustable parameter, and can be used to accommodate a range of target distributions across the CCDs.  For example, a perfectly even distribution of targets could be obtained by dividing the limits assigned per CTL (Table~\ref{tab:ctls})  by 16 and setting the parameter to this value for all CCDs. On the other hand, the scientific priorities can completely determine the target selection by setting the limits on each CCD to 2\,000  targets (the maximum allowed by ground systems). Intermediate cases that balance these two options can be obtained by setting the limits for each CCD between these two extremes.  The limits for individual CCDs also do not need to have the same value for a given CTL, and so different CCDs/cameras can be more highly weighted with respect to the scientific priorities versus technical constraints.  
 
No CCD limits were applied to the bright, astero, and GI CTLs---owing to their small size, it is more important to select as many high priority targets as possible rather than enforce an even distribution across CCDs.  The exo CTL, owing to its large number of target slots, can absorb the work of dividing targets across the many CCDs for systematic error correction.  However,  we give extra weight to the scientific priorities of exo CTL targets in Camera~4, because these targets will have long baseline light curves and therefore additional scientific value.  We assign 1\,400 exo targets  to each of the 4 CCDs in Camera~4 from the exo CTL, with 650 assigned to each of the other 12 CCDs in Cameras~1, 2, and 3.  The DDT list is assigned a maximum of 500 targets per CCD.  Although this limit suggests no particular preference per CCD for the DDT CTL, the total number is well beyond the allotment in Table~\ref{tab:ctls} and allows DDT targets to be distributed mainly based on scientific priority.
 
The available slots for Camera~4 have usually filled up by the time the second pass of the exo CTL is implemented.  Thus, many higher priority targets in this region of the sky are skipped in order to add targets to the other three cameras.  In some cases, the slots in Camera~4 are exhausted during the GI target selection---this depends specifically on the observing sector, patch of sky, and available targets and their priorities in other cameras.
 
TESS is usually pointed at +54 degrees ecliptic latitude so that the center of the field of view of Camera~4 is near the ecliptic pole.  However, in six sectors of year 2, TESS was pointed at +85 degrees in ecliptic latitude to avoid  high backgrounds caused by earthshine in Camera~1 (Sectors~14--16 and 24--26).  In these cases, Cameras~2 and 3 straddle the north ecliptic pole, and we adjusted the weighting of targets across CCDs/cameras and CTLs to account for this.  All CCDs in Camera~3 were allotted 1\,150 exo CTL slots, along with the  two CCDs in Camera~2 nearest the ecliptic pole.  The other CCDs were assigned 650 exo CTL slots.  No change was made for the number of targets per CCD for the other CTLs.

\floattable
\begin{table}[htb]
\begin{center}
\caption{Target Selection Tests}
\label{tab:target_tests}
\begin{tabular}{p{0.02\columnwidth} p{0.5\columnwidth}  p{0.3\columnwidth} }
\hline
 &\multicolumn{1}{c}{Test} & \multicolumn{1}{c}{Action if Yes}\\
\hline
1 & Are any pixels associated with this target outside of the science imaging region or overlapping with bleed trails of saturated stars?  & Go to next target.\\
2 & Would adding the target exceed the total number of targets for this CCD? & Ignore this CCD, go to next target.\\
3 & Would adding the target exceed the total number of targets for this CTL? & Go to next CTL.\\
4 & Would adding the target exceed the total number of targets for this sector? & Exit and complete target selection.\\
\hline
\end{tabular}\\
\end{center}
\end{table}

\subsection{Selection of Targets per CTL\label{sec:individual_targets}}
For each CTL, the POC adds targets with the following procedure.

  First, we filter each CTL to identify which targets land on the imaging regions of the CCDs (see \ref{sec:pixel_selection}).  Next, we sort the targets by decreasing priority.  For the highest priority target, we then check if adding its associated pixels violates any technical requirements---see \S\ref{sec:pixel_selection} for a description of how pixel subarrays are assigned for each target and Table~\ref{tab:target_tests} for a full list of tests.  If the target passes all tests, we add it to the target list, update the pixel table to track the associated subarray pixels, and write files that command the instrument to save these pixels at 2-minute cadence.  We then repeat this process for the next-highest priority target and iterate until reaching the maximum number of targets per CTL, or exhausting the observable targets from the CTL.  If a target lands on a CCD which is already full for a given CTL (\S\ref{sec:ccd_dist}), that target is skipped.

Storage space in the SSR onboard the spacecraft also implies a limit on the total number of pixels that can be collected at 2-minute cadence, and a parameter is available to limit the total number of pixels associated with 2-minute targets.  However, limiting the total number of pixels for 2-minute targets would lead to a variable number of selected targets per sector.  Instead, we use the total number of targets to limits the 2-minute target selection---20\,000 targets have been found to produce a data volume close to (but always less than) the total storage space reserved for 2-minute cadence data.

The total number of targets per CCD is tracked.  Once the maximum number of targets on a given CCD is reached, candidate targets that land on that CCD are skipped.

The total number of targets per CTL is tracked.  Once the slots allotted to that CTL are filled, the remaining targets from that CTL are ignored and the selection process moves on to the next CTL.

The total number of targets per sector is tracked.  Once the maximum number of targets is reached, the target selection loop exits, logs of the targets and their CTLs of origin are saved, and files are produced for tracking the target list/pixel table and commanding the spacecraft.  If there are still slots remaining after searching all six CTLs, we return to the exo CTL to fill out any remaining space.

After completing target selection, the final target list is checked against each CTL to find the total number of targets observed per CTL.  This number is always larger than the allocated slots per CTL, because targets that are listed at low priority in one CTL are often selected at high priority by another CTL.

\subsection{Selection of Pixel Subarrays per Target\label{sec:pixel_selection}}
The subarray pixels associated with each selected target must contain the optimal photometric aperture and enough background pixels to robustly estimate the sky level. For the majority of targets, an 11$\times$11 pixel box is sufficient.  For saturated stars, additional rows are collected to fully capture the bleed trails.  

The locations of stars in the image are calculated with the TESS focal plane geometry (FPG) model---a full description of the FPG model will be presented in future work.  The FPG model is known to be accurate to a few hundredths of a TESS pixel.  When calculating the expected position of targets in the TESS field of view, celestial coordinates are corrected for proper motion and velocity aberration.  Once the location of a target is identified, it is trivial to define the corresponding 11x11 pixel subarray.  

For saturated stars, we collect additional rows proportional to the expected flux beyond a "saturation magnitude."  The saturation magnitude is defined as the magnitude of a star whose integrated flux would saturate a single TESS pixel in a single two-second exposure.  The TESS pixel full-well depth is approximately 200\,000 photoelectrons per pixel, with the exact number varying depending on the CCD output node (see \citealt{TIH}, Figure~4.2).  Adopting 200\,000 photoelectrons per two seconds as the saturation flux results in a saturation magnitude of ${\rm T_{mag}}=$ 7.5.  We calculate the total number of additional rows by scaling the expected flux of the source relative to the 7.5 magnitude saturation magnitude.  As an example, a 2.5 magnitude target has 100 times the flux of a 7.5 magnitude target, and so we expect to require 100 extra rows to collect all the saturated pixels for the brighter source.  We allocate half of the extra rows above and half below the original 11x11 pixel subarray.  If the calculation results in an odd number of extra rows, we always add one additional row so that the extra rows can be distributed evenly above and below the original pixel subarray.

This procedure results in subarrays that grow somewhat faster than the observed bleed trails of bright stars.  The faster growth is caused by pointing jitter and the finite width of the instrument's PSF, both of which spread light from a point source across multiple pixels and effectively diminish the amount of flux individual pixels receive relative to the saturation magnitude.  It is preferable to have a buffer of pixels around the bleed trails in the subarrays of saturated stars for users interested in photometry of these stars, and so we have not found it necessary to tune the bleed trail growth calculation further. The procedure works well unless the magnitudes of the bright stars are underestimated in the TIC, or multiple bright stars land close enough to each other (within a few pixels)  that the combined light saturates the central pixel faster than expected.  Although uncommon, several examples have been collected---see the TESS Data Release Notes for specific examples.\footnote{\url{https://archive.stsci.edu/tess/tess_drn.html}}

Apart from extra rows collected for bleed trails, all pixels associated with a target must land in the science imaging region of the target CCD.  If a target identified as "on silicon" is too close to the edge of a CCD image (within 5 pixels) it is not added to the final target list and the target selection algorithm continues to the next-highest priority target.  The target is also skipped over if any of its pixels are within three columns of a saturated star and the associated bleed rows, where the bleed rows are calculated in the same way as above.

There is a small amount of overlap, about 10 pixels, between the fields of view of each camera.  In cases where a target lands in this region, the target's pixel subarray is assigned to the camera farthest from the ecliptic pole in order to save target slots for higher priority targets.  Because we require the full 11x11 subarray to be observable, this only happens to targets that land on rows 6--10 of a given CCD.

Each CTL may also request a different shape of the subarray using a keyword provided in the CTL.  In practice, some GI programs have requested  25x25 pixel apertures.  In cases where different CTLs request different pixels (e.g., for larger apertures), the superset of pixels is selected for download.

\subsection{PPA Star Selection\label{sec:ppa_selection}}
The POC generates a list of PPA stars for each sector, which are used to monitor the instrument performance and derive WCSs for each subarray and FFI.  As a reminder, PPA stars must be bright but unsaturated, show little astrophysical variability, be reasonably well isolated in the TESS images (to avoid crowding and possible bias in their position measurements), and be evenly spread out across the TESS fields of view.  These requirements are often at odds with each other, in particular when a particularly crowded region of the sky (such as an open cluster or the Galactic plane) would cause a large gap in the coverage of PPA stars across a TESS image. 

The POC developed an algorithm to perform the PPA selection in a way that balances these requirements, and includes adjustable parameters that can be used to override the default PPA selection for especially difficult fields. The procedure starts by generating a list of stars between 8th and 10th magnitude that land in the TESS image.  The stars must be 5 pixels away from the image edges, as well as 5 pixels away from the edges of CCD output nodes (see \citealt{TIH}, Figure~4.2).  We then remove stars that are known variables based on the Variable Star Index \citep{Watson2006}, and we remove stars that land near the bleed trails of bright saturated stars (see \S\ref{sec:pixel_selection}).  Next, we identify and remove stars with near neighbors using the following criteria:  there must be no star brighter than ${\rm T_{mag}} = {\rm T_ {PPA}} + 2$ within 5 pixels of the PPA star candidate, where ${\rm T_ {PPA}}$ is the ${\rm T_{mag}}$ of the PPA star candidate, and no star brighter than ${\rm T_{mag}} = {\rm T_ {PPA}} + 5$ within 3 pixels.  Stars within 16 pixels of a 6th magnitude star or brighter are also removed, because light from the nearby bright companion can affect the background and centroid estimates of the PPA star.  These pixel and magnitude thresholds are adjustable parameters, and can be relaxed if the field of view is exceptionally crowded and few or no viable PPA stars are identified (as occurs near the Galactic center, for example).  There is a minimum of 100 PPA stars required for every sector---if fewer stars than this are found, target selection is rerun with adjusted crowding parameters.

Usually several hundred suitable PPA stars per CCD can be found with this method.  However, a maximum of 120 PPA stars per CCD is imposed because only 100 PPA star centroids are needed to obtain accurate astrometry and WCSs.  The final step of the POC algorithm is to select the subset of 120 stars that most evenly cover the CCD image.  Firstly, if fewer than 120 stars are available (but more than 100), a better set of stars cannot be found for this field of view and the  stars are selected as PPAs.  If more than 120 stars are available, the POC finds pairs of candidate PPA stars and orders the pairs by increasing separation.  For each pair in order of increasing separation, the faintest star is removed until 120 PPA stars remain.  This results in the maximally spaced set of good PPA stars available on each CCD.  We always check the spatial distribution of PPA stars and their magnitudes before uploading the target list tables to the spacecraft---if there are problems (for example, an exceptionally large gap between PPA stars in some portion of the image), we rerun the target selection with adjusted crowding parameters until a suitable distribution of PPA stars is found.

\subsection{Target Selection in the TESS Extended Mission\label{sec:em_selection}}
\floattable
\begin{deluxetable}{cccr}
\tablewidth{0pt}
\tablecaption{TESS Extended Mission Candidate Target Lists\label{tab:em_ctls}}
\tablehead{
\colhead{Order} &\colhead{Name} & \colhead{Origin}  &\colhead{N Slots}
 }
 \startdata
 20-second Cadence\\
 \hline
1 & ppa & POC   &400\\
2 & GI  & GI Office & 540\\
3& DDT & PI/POC&61 \\
\hline
 2-minute Cadence\\
 \hline	
1 & ppa & POC   &1\,920\\
2 & bright & POC \\
3 & 20-second & POC &  1\,000\\
4 & GI  & GI Office  & 12\,001\\
5 & DDT & PI/POC &1\,501 \\
6 & exo & TSWG/POC &  1\,501\\
\enddata
\end{deluxetable}

In the first TESS extended mission, starting 2020 July 4, TESS began selecting a limited number of targets for observations at 20-second cadence.  The POC uses the same target selection algorithm described above for the 20-second cadence targets, but with parameter changes  driven by  tighter data margins in the extended mission (caused by 20-second cadence data and 10-minute cadence FFIs).  These changes are described below and summarized in Table~\ref{tab:em_ctls}.
\begin{enumerate}
\item Only 1\,000 total targets per sector are observed at 20-second cadence, with no more than 400 targets observed per CCD.
\item For 20-second cadence observations, 25 PPA stars per CCD are selected instead of 120 PPA stars for 2-minute observations.  The smaller number of stars degrades the accuracy and precision of the WCSs for 20-second target data, but the trade-off is acceptable given the limited number of 20-second target slots.
\item 20-second cadence CTLs are produced by the GI office and DDT program.  The POC selects targets from the GI CTL first, followed by the DDT CTL.  If any remaining target slots are available, the GI CTL is searched a second time.
\item On the first pass of the 20-second GI CTL target selection, 540 targets are selected.  A maximum of 40 targets per CCD is imposed on Cameras~1 and 2, a maximum of 80 targets per CCD on Camera~3, and a maximum of 160 targets per CCD on Camera~4.  The DDT targets are limited to 60 total, with a maximum of 10 targets per CCD.  These parameters will be adjusted in Year~4 observations for ecliptic pointings and pointing adjustments that avoid scattered light in Camera~1.
\end{enumerate}

The 2-minute target selection changed in the extended mission in the following ways:
\begin{enumerate}
\item The POC adds a CTL listing the 20-second cadence targets to the 2-minute target selection, so that all targets observed at 20-second cadence are also observed at 2-minute cadence.  This is done in order to monitor the pipeline performance, particularly with respect to cosmic ray removal in 20-second cadence data and noise properties of the light curves.
\item The POC selects 1\,501 targets from the exo CTL and 12\,001 targets from the GI CTL.  The order of selection also changes: ppa, bright, 20-second cadence, GI, DDT, exo.  There is no astero CTL in the extended mission separate from targets proposed through the GI program.  If any target slots remain, the GI CTL is searched a second time, and if there are still remaining target slots, the exo CTL is searched a second time.
\item The GI targets are subjected to the same process described above for of evenly distributing targets over  CCDs for systematic error correction.  There is a maximum of 660 targets per CCD in Cameras~1--3, and 1060 targets per CCD in Camera 4.  No other CTL has per CCD limits applied.
\item The exo CTL is reprioritized in the extended mission, favoring stars promoted to TESS Objects of Interest during the primary mission \citep{Guerrero2021}.  Stars that landed in gaps between the TESS CCDs in the primary mission observations are also added, but at lower priority.  After these two lists, stars with potential signs of transits in the primary mission, but at low signal-to-noise ratio, are added at the lowest priority.  The majority of these targets are likely false positives due to measurement noise, but the targets are added because additional observations in the extended mission might improve the transit search for very small planets.  
\end{enumerate}

\section{Results\label{sec:results}}
In this section, we assess the POC target selection over the first two years of the mission.  

\subsection{Total targets observed in Year~1 and 2}

\begin{figure}
  \centering
  \includegraphics[width=0.48\textwidth]{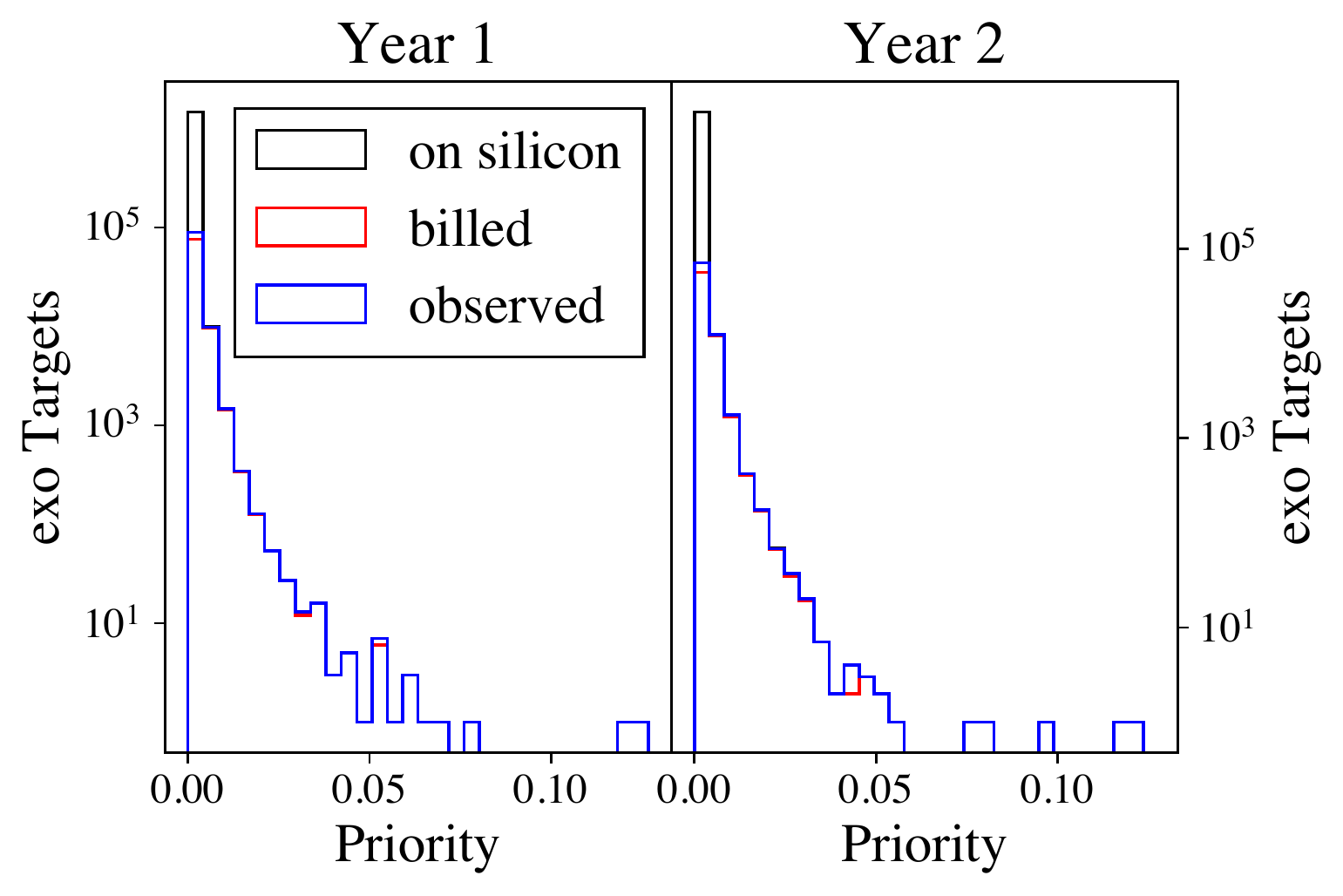}
  \caption{Histograms of priority of exo CTL targets observed in the TESS primary mission.  The labels ("on silicon," "billed," and "observed") are the same as in Figure~\ref{fig:example_sector} and are defined in \S\ref{sec:merging_ctls}.  The year~1 and year~2 targets are separated because they were drawn from different CTLs tied to different versions of the TIC (v07 in year~1 and v08 in year~2, see \citealt{Stassun2019}).\label{fig:exo_hist}}
\end{figure}

\begin{figure*}
  \centering
  \includegraphics[width=\textwidth]{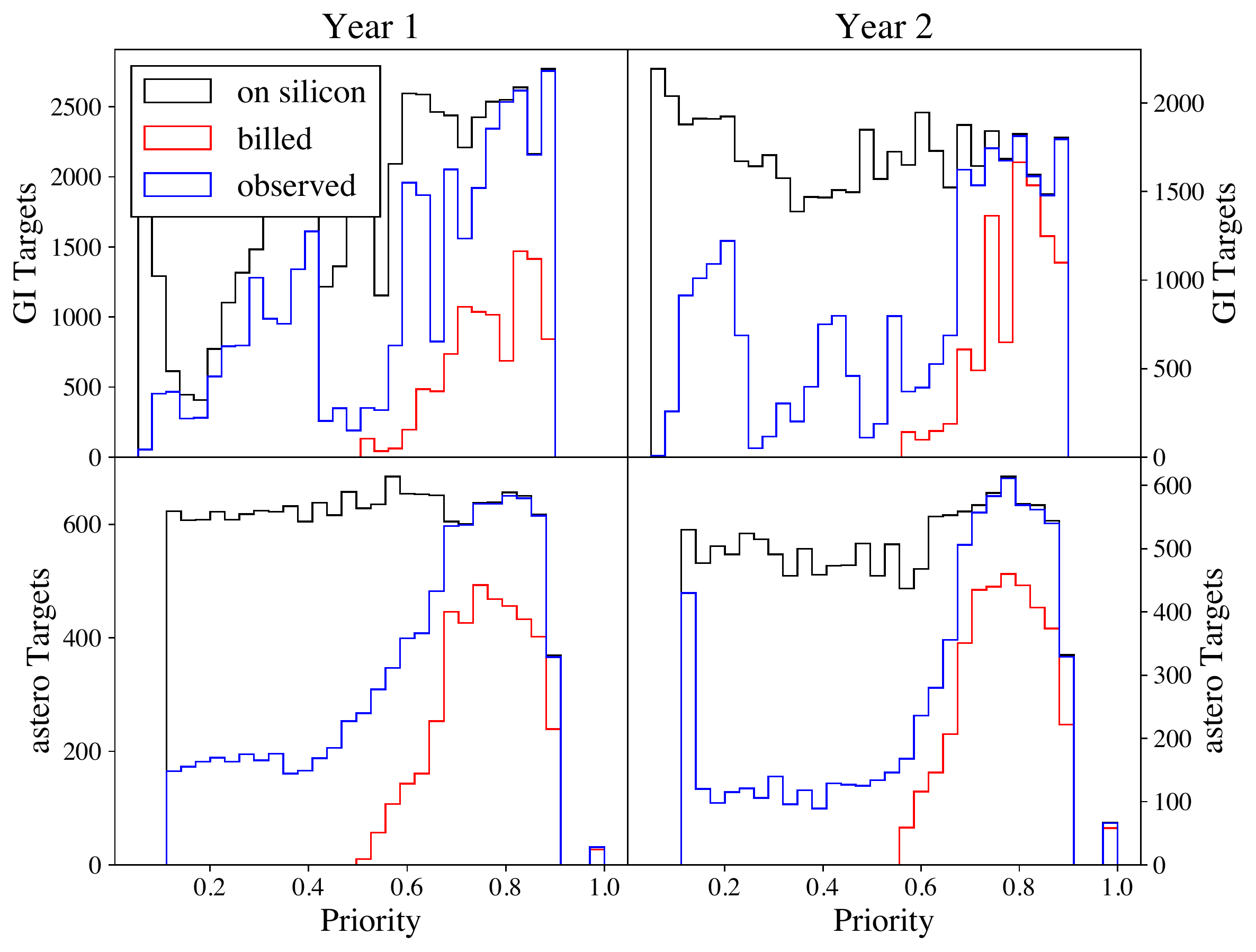}
  \caption{Same as Figure~\ref{fig:exo_hist}, but for the GI and astero targets.  The differences between the blue and red histograms is caused by overlap between the various CTLs, and shows that many more targets were observed than initially allocated to the GI and astero programs (see Table~\ref{tab:ctls} and \S\ref{sec:merging_ctls}).\label{fig:astero_GI_hist}}
\end{figure*}

\begin{figure*}
\centering
\includegraphics[width=\textwidth]{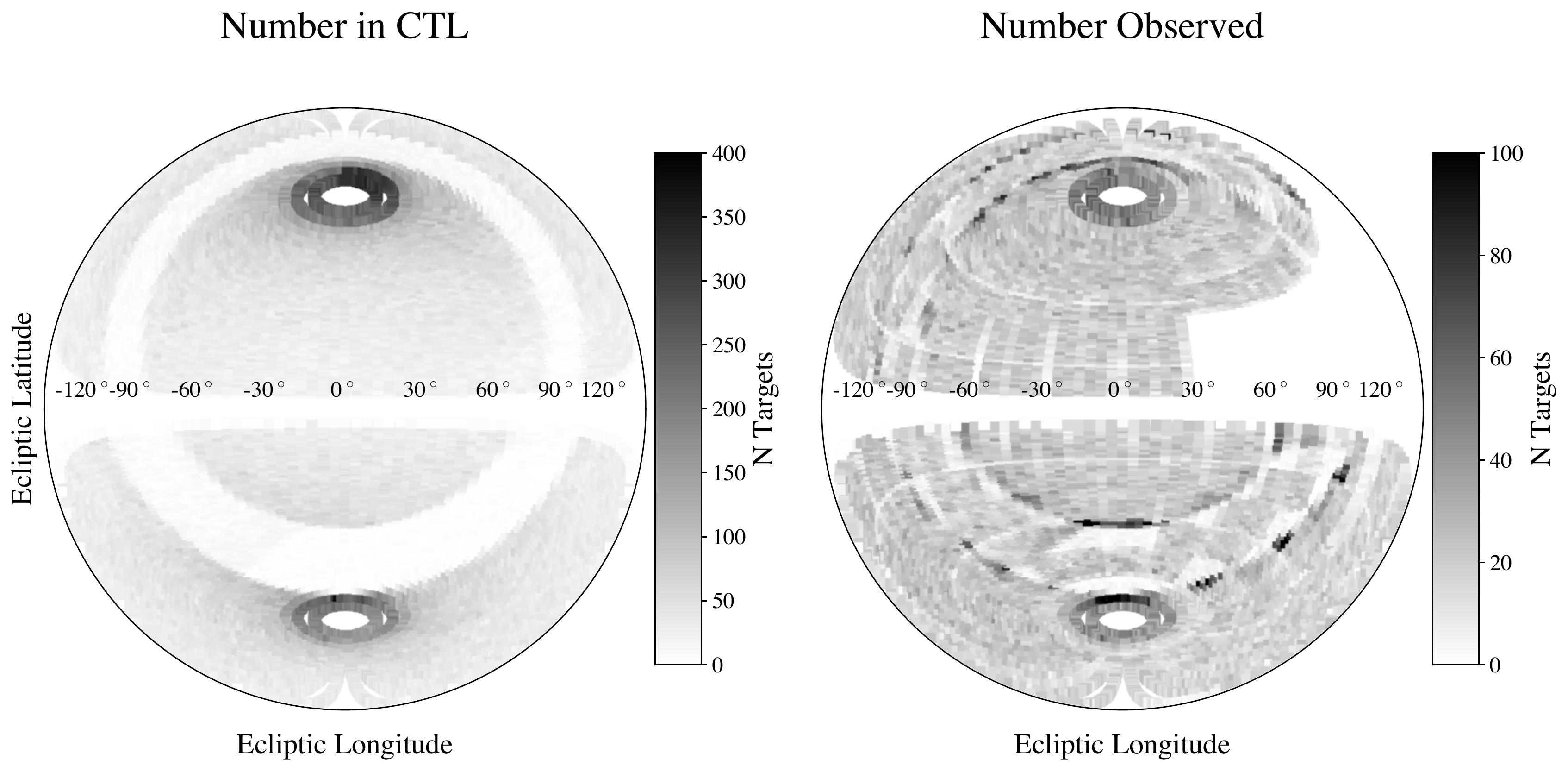}
\caption{The left panel shows the sky distribution in ecliptic coordinates of the top 200 000 stars by priority in the exo CTL used for target selection in the TESS primary mission. The bins subtend equal solid angles, and the white-space at the poles is an artifact of the projection.  Note that the southern and northern hemispheres were constructed from different CTLs (v07 and v08, respectively).  The curvilinear band is the plane of the Galaxy, which was preferentially avoided because the amount of crowding/contamination expected in the large TESS pixels. The right panel shows the targets that were actually observed for 2-minute observations.  The difference between the two panels shows how technical constraints of the mission impact the sky distribution of observed targets.
\label{fig:sky_dist}
}
\end{figure*}

The mission imposed a requirement that over 200\,000 targets be observed at 2-minute cadence in the two-year primary mission.  For year~1, 128\,292 unique targets were observed, and, for year~2, 104\,473 unique targets were observed.

Figure~\ref{fig:exo_hist} shows the histogram of priorities of selected targets for the exo CTL.  The targets from year~1 (Sectors~1--13) and year~2 (Sectors~14--26) are shown separately because the version of the TIC and CTL changed in Sector~14, along with associated priorities in the CTLs.  This change corresponds to TIC~8, which is based on Gaia DR2 rather than 2MASS---details are given in \citet{Stassun2019}.  Figure~\ref{fig:exo_hist} also shows the subsets of "billed" and "observed" targets  compared to the parent population of observable targets ("on silicon").  Almost all of the higher priority targets were selected for 2-minute observations.  There is little difference between the billed and observed targets because the exo CTL targets are selected early in the target selection process to the benefit of the astero and GI CTLs (see \S\ref{sec:merging_ctls}).

Figure~\ref{fig:astero_GI_hist} shows the histograms of priorities of targets for the astero and GI CTLs.  As for Figure~\ref{fig:exo_hist}, the year~1 and year~2 targets are shown separately, as well as the on silicon, billed, and observed subsets of targets.  The astero and GI CTLs are allocated a small number of targets relative to the exo CTL, but many more targets were observed because of the CTL merging procedure described in \S\ref{sec:merging_ctls}.  The overlap between the various CTLs also adds additional targets at lower priority.  The differences between the red and blue histograms in Figure~\ref{fig:astero_GI_hist}  serves as a simple metric for the success of the target selection algorithm, because a larger difference indicates additional targets from each CTL that bolster the scientific return of the mission.  

The mission also imposed a requirement that $>$5\,000 targets be observed for more than 250 days in each hemisphere.  A total of 6\,010 such targets were observed in year~1, and 7\,162 such targets in year~2.  Prior to launch, we experimented with forcing stars observed in previous sectors to be reselected in subsequent sectors, regardless of priority.  This resulted in relatively small changes in the number of targets observed for more than 250 days, only about 300 more targets than were obtained in a single hemisphere.  We therefore did not force stars selected in early sectors to be re-observed in subsequent sectors.

\subsection{Sky Distribution\label{sec:sky_dist}}

Figure~\ref{fig:sky_dist} shows the sky distributions of targets from the exo CTLs and those actually observed on the sky.  The left panel shows the density of the highest priority 200\,000 targets in each hemisphere.  The right panel shows the density of exo targets selected for observation.  Note that different versions of the TIC and CTLs are displayed for the southern (v7) and northern (v8) ecliptic hemispheres.

While there is a smooth distribution with decreasing density from the poles to the ecliptic plane in the CTL histogram (left panel), the dominant features of the selected target histogram (right panel) are the discrete boundaries caused by gaps between the CCDs.  The density of targets on individual CCDs is mostly flat, and any gradient between the poles and ecliptic is much weaker than in the parent CTL population.  The main exceptions are for CCDs that contain the plane of the Galaxy.  Crowding due to the high density of stars in the Galactic plane causes targets in this area of the sky to have significantly lower priorities than the rest of the CTL.  A single CCD that contains the Galaxy will therefore preferentially add targets away from the Galactic plane, resulting in a clumpy distribution of targets around the edges of these CCDs. 

The differences between the sky distributions of CTL targets and selected targets in Figure~\ref{fig:sky_dist} are caused by the technical requirements of the mission, particularly the requirement that targets be roughly evenly distributed over all TESS CCDs.  This requirement is applied to improve the light curve systematic error correction in the TESS mission pipeline (PDC).   However, the complicated interaction of the technical requirements and CTL priorities results in an inhomogeneous sample of target stars.  In general, the properties of the CTL catalog cannot be used to infer the properties of the sample of stars actually observed by TESS.  Therefore, we suggest that researchers use carefully tailored subsamples of TESS targets when studying the statistical properties of TESS planet hosts.  A detailed comparison of the stellar properties of the CTL and the stars actually observed for 2-minute observations by TESS will be presented in future work (Pepper et al., in preparation).

\section{Conclusion\label{sec:conclusion}}
We have described technical limits that constrain target selection for TESS observations and the algorithms by which the POC selects targets from the CTLs.  Our two main results are 

\begin{enumerate}
\item The POC algorithms result in factors of 2.1 to 3.4 times as many observed targets  from a given CTL as target slots allocated to that CTL, representing an  improved scientific return compared to baseline expectations.
\item  Technical limitations result in a different population of observed stars compared to the underlying properties of the input CTLs.  We caution researchers exploring statistical analyses of TESS planet-host stars that the population of observed targets cannot be inferred from the properties of the input CTLs. A detailed comparison of the stellar properties of both samples will be presented by Pepper et al., in prep.
\end{enumerate}

  We have also described changes and updates to the target selection procedures for the TESS extended mission, including the new 20-second cadence data mode.  These parameters will be in place through the end of the 26 month Extended Mission 1 (2022 September).  However, these parameters ensure that the POC algorithms are very flexible, and can be adjusted to whatever needs the mission may require for programmatic changes in the future.

  \acknowledgements
    Funding for the TESS mission is provided by NASA's Science Mission Directorate.  Resources supporting this work were provided by the NASA High-End Computing (HEC) Program through the NASA Advanced Supercomputing (NAS) Division at Ames Research Center for the production of the SPOC data products.  
    
The TESS mission appreciates contributions to development and/or operations from P.~R.~McCullough,  G.~Laughlin, and B.~Shiao.
    
GB gratefully acknowledges the opportunity of spending a sabbatical year at the Institute for Advanced Study.    Part of this research was carried out at the Jet Propulsion Laboratory, California Institute of Technology, under a contract with the National Aeronautics and Space Administration (NASA).  Funding for the Stellar Astrophysics Centre is provided by The Danish National Research Foundation (Grant DNRF106).  Work by B.S.G. was funded by the Thomas Jefferson Chair for Discovery and Space Exploration. 

\section*{Appendix A \label{app:sector_targs}}

Table~\ref{tab:sector_targets} compares the number of billed and observed targets for each CTL.  These data are represented graphically in Figure~\ref{fig:sector_targs}.

\begin{deluxetable*}{r|r|r|rrr|rrr|rrr|rrr}\tablewidth{0pt}\tablecaption{Number of Targets for each CTL by Sector\label{tab:sector_targets}}\tablehead{ \colhead{\phantom{Sector}}&\colhead{ppa} &\colhead{bright} &\multicolumn{3}{c}{exo} &\multicolumn{3}{c}{astero} &\multicolumn{3}{c}{GI} &\multicolumn{3}{c}{DDT}} \startdata Sector&billed &billed & billed &observed &ratio &billed &observed &ratio &billed &observed &ratio &billed &observed &ratio \\ \hline
       1 &     1920 &      547 &    11180 &    13008 &     1.16 &      751 &     1697 &     2.26 &     1501 &     5303 &     3.53 &        0 &        0 &     0.00\\
       2 &     1920 &      513 &    10860 &    13230 &     1.22 &      751 &     1557 &     2.07 &     1501 &     5327 &     3.55 &       26 &      176 &     6.77\\
       3 &     1920 &      475 &    10860 &    13383 &     1.23 &      751 &     1520 &     2.02 &     1501 &     5314 &     3.54 &       28 &      138 &     4.93\\
       4 &     1920 &      451 &    15000 &    17213 &     1.15 &      751 &     1647 &     2.19 &     1501 &     5763 &     3.84 &       71 &      207 &     2.92\\
       5 &     1920 &      569 &    15000 &    17089 &     1.14 &      751 &     1733 &     2.31 &     1501 &     5879 &     3.92 &      142 &      865 &     6.09\\
       6 &     1920 &      748 &    13400 &    17003 &     1.27 &      751 &     1612 &     2.15 &     1501 &     5290 &     3.52 &      139 &      722 &     5.19\\
       7 &     1920 &      797 &    13400 &    17046 &     1.27 &      751 &     1569 &     2.09 &     1501 &     5064 &     3.37 &       95 &      633 &     6.66\\
       8 &     1920 &      718 &    15000 &    17040 &     1.14 &      751 &     1560 &     2.08 &     1501 &     5381 &     3.58 &      110 &      816 &     7.42\\
       9 &     1920 &      668 &    13400 &    16653 &     1.24 &      751 &     1504 &     2.00 &     1501 &     5286 &     3.52 &     1359 &     2125 &     1.56\\
      10 &     1920 &      753 &    13400 &    16653 &     1.24 &      751 &     1545 &     2.06 &     1501 &     5465 &     3.64 &     1204 &     2018 &     1.68\\
      11 &     1919 &      849 &    13400 &    16577 &     1.24 &      751 &     1488 &     1.98 &     1501 &     5480 &     3.65 &     1141 &     1962 &     1.72\\
      12 &     1920 &      823 &    13400 &    16758 &     1.25 &      751 &     1602 &     2.13 &     1501 &     5191 &     3.46 &      936 &     1753 &     1.87\\
      13 &     1920 &      647 &    13400 &    16696 &     1.25 &      751 &     1752 &     2.33 &     1501 &     5245 &     3.49 &     1345 &     2172 &     1.61\\
      14 &     1920 &      804 &    13400 &    16767 &     1.25 &      751 &     1583 &     2.11 &     1501 &     4843 &     3.23 &      127 &      640 &     5.04\\
      15 &     1919 &      776 &    13400 &    16813 &     1.25 &      751 &     1586 &     2.11 &     1501 &     4882 &     3.25 &      131 &      610 &     4.66\\
      16 &     1920 &      755 &    13400 &    16811 &     1.25 &      751 &     1564 &     2.08 &     1501 &     4822 &     3.21 &      113 &      616 &     5.45\\
      17 &     1920 &      816 &    13400 &    16684 &     1.25 &      751 &     1532 &     2.04 &     1501 &     4876 &     3.25 &       97 &      552 &     5.69\\
      18 &     1920 &      810 &    13400 &    16723 &     1.25 &      751 &     1556 &     2.07 &     1501 &     4944 &     3.29 &      107 &      571 &     5.34\\
      19 &     1920 &      749 &    13400 &    16917 &     1.26 &      751 &     1480 &     1.97 &     1501 &     4793 &     3.19 &      100 &      563 &     5.63\\
      20 &     1920 &      629 &    13400 &    17131 &     1.28 &      751 &     1553 &     2.07 &     1501 &     4789 &     3.19 &       89 &      562 &     6.31\\
      21 &     1920 &      536 &    13400 &    17332 &     1.29 &      751 &     1543 &     2.05 &     1501 &     4813 &     3.21 &       93 &      567 &     6.10\\
      22 &     1920 &      495 &    13400 &    17395 &     1.30 &      751 &     1591 &     2.12 &     1501 &     4873 &     3.25 &       93 &      553 &     5.95\\
      23 &     1920 &      521 &    13400 &    17408 &     1.30 &      751 &     1569 &     2.09 &     1501 &     4864 &     3.24 &       98 &      564 &     5.76\\
      24 &     1920 &      701 &    13400 &    17916 &     1.34 &      751 &     1526 &     2.03 &     1501 &     5245 &     3.49 &      127 &      677 &     5.33\\
      25 &     1920 &      678 &    13400 &    16942 &     1.26 &      751 &     1513 &     2.01 &     1501 &     5147 &     3.43 &      120 &      672 &     5.60\\
      26 &     1919 &      711 &    13400 &    16877 &     1.26 &      751 &     1509 &     2.01 &     1501 &     5128 &     3.42 &      140 &      693 &     4.95\\
\hline  Average &     1920 &      675 &    13304 &    16541 &     1.24 &      751 &     1573 &     2.09 &     1501 &     5154 &     3.43 &      309 &      824 &     2.67\\
\enddata\end{deluxetable*}

\end{document}